\documentclass[epj]{svjour}

\RequirePackage{graphicx}
\usepackage{graphicx}
\usepackage{dcolumn}
\usepackage{bm}
\usepackage{amsmath}
\usepackage{amssymb}

\def\be{\begin{equation}}
\def\ee{\end{equation}}
\def\bq{\begin{eqnarray}}
\def\eq{\end{eqnarray}}
\journalname{Eur. Phys. J. A}
\begin{document}
\title{Strangeness thermodynamic instabilities in hot and dense nuclear matter}
\author{A. Lavagno and D. Pigato}
\institute{Department of Applied Science and Technology, Politecnico di Torino, I-10129 Torino, Italy and \\
Istituto Nazionale di Fisica Nucleare (INFN), Sezione di Torino, I-10126 Torino, Italy}
\date{Received: date / Revised version: date}
\abstract{We explore the presence of thermodynamic instabilities and, con\-se\-quen\-tly, the realization of a pure hadronic phase transition in the hot and finite baryon density nuclear matter. The analysis is performed by means of an effective relativistic mean-field model with the inclusion of hyperons, $\Delta$-isobars, and the lightest pseudoscalar and vector meson degrees of freedom. The Gibbs conditions on the global conservation of baryon number and zero net strangeness in symmetric nuclear matter are required. Similarly to the liquid-gas phase transition, we show that a phase transition, characterized by mechanical instabilities (due to fluctuations on the baryon number) and chemical-diffusive instabilities (due to fluctuations on the strangeness number), can take place for a finite range of $\Delta$-meson coupling constants, compatible with different experimental constraints. The hadronic phase transition, which presents similar features to the quark-hadron phase transition, is characterized by different strangeness content during the mixed phase and, consequently, by a sensible variation of the strange anti-particle to particle ratios.
} 
\maketitle

\section{Introduction}
One of the major challenges in the high energy heavy-ion collisions is a detailed study of the nuclear equation of state (EOS) at different regimes of baryon chemical potential and temperature, with the investigation of possible phase transition phenomena during the collisions \cite{satz2018}.

At high temperature regime, various QCD inspired theoretical models indicate a region with a rapid cross-over of thermodynamic observable and a formation of a critical endpoint, beyond which the system shows a first order phase transition from confined to deconfined matter \cite{raja2020,albe2020,blaschke2018,biro2018,wilk2019}. The existence and the location of such phase transition at finite baryon chemical potential is still a matter of debate and can be in principle detected in the planned high-energy compressed nuclear matter experiments \cite{cmb2021,jparc2021,shine2020,bleicher2020,star2017,blaschke2016}.

At low temperatures and subnuclear densities, a li\-quid-gas type of phase transition was predicted and observed in nuclear multifragmentation experiments at inter\-me\-dia\-te-energy \cite{kapusta1984,pocho1995,das2018,mallik2021}.
Because nuclei are made of protons and neutrons with two conserved charges (baryon number and electric charge), such a phase transition is continuous (rather than discontinuous as in the one-com\-po\-nent system) and, consequently, for a binary system, the instabilities in the mixed liquid-gas phase arise from fluctuations in the baryon density and in the proton concentration \cite{muller1995,mekjian2003,muller1997,physica2013}.

Recently, the study of a nuclear liquid-gas phase transition has been extended to the strangeness sector at low temperature regime, below and above the nuclear saturation density, in order to examine the occurrence of phase transitions and thermodynamic instabilities in presence of the hyperon degrees of freedom \cite{wang2004,yang2006,das2015,gulminelli2016,das2017}. The relevance of strangeness instabilities has been also studied in the context of dense $\beta$-stable neutron star and supernova matter \cite{gulminelli2017}.

In relativistic heavy-ion collisions, besides hyperons, a state of high density resonance $\Delta(1232)$-isobar matter may be formed. Transport model calculations and experimental results indicate that an excited state of baryonic matter is dominated by $\Delta$-resonance at the energy from AGS to RHIC \cite{hofmann1995,lavagno2013,stoecker2018,lenske2018,takeda2018}.
In addition, it has been pointed out that the existence of $\Delta$-isobars can be very relevant also in the core of neutron stars \cite{prd2014,voskresensky2017,physica2019,weber2018,weber2020,raduta2020,raduta2021}.

In this context, is important to remember that the recent discovery of massive neutron stars and different astrophysical observations, mainly related to neutron star mergers with electromagnetic and gravitational  wave signals, put strong constraints on the EOS of dense baryonic matter, which must be rather stiff to support a large mass against gravitational collapse \cite{bednarek,fujimoto,gorda}.  Moreover, the existence of massive compact stars ($M\ge 2.1 M_\odot$) implies that the speed of sound (strictly related to the stiffness of the EOS) exceeds the conformal limit ($c_s^2=1/3$, in units of the speed of light) in the scenario of one family hadronic EOS \cite{bedaque,annala,rezzolla2022}. Therefore, it would be some physical mechanism for which the speed of sound as a function of density should increase to values significantly larger than the conformal bound, with at least one local maximum, and it should decrease to asymptotically reach the conformal limit, in agreement with the pQCD calculations \cite{hoyos,reddy} \footnote{For the sake of completeness, we remember that lattice QCD calculations had clearly established the speed of sound at finite temperature and zero density matter is always below the conformal limit \cite{karsch}.}.
On the other hand, the appearance in the EOS of hyperons and $\Delta$-isobars implies a remarkable softening of the EOS at high density with resulting a significant reduction of the achievable maximum mass. As discussed in Refs. \cite{prd2014,prc2014}, this problem could be overcome in the scenario of two coexisting families of compact stars: hadronic stars, whose EOS is soft (like the one adopted in the present investigation), can be very compact with small radii and with maximum masses of about $1.5 M_\odot$, while massive strange quark stars, whose EOS is stiff, with masses greater than 2 $M_\odot$ \cite{berezhiani2003,epja2016}.

The scenario of two-family compact stars implies that hadronic matter is metastable and decays into strange quark matter, by assuming the Bodmer-Witten hypothesis \cite{witten}. The condition for a nucleation conversion from beta-stable hadronic stars to quark stars are related to a critical amount of net strangeness (or hyperons fraction) that is present in the cold beta-stable hadronic star \cite{depietri}. Recent investigations have shown that in this latter scenario, strange quark stars can reach very massive conditions (larger also than 2.5 $M_\odot$, achievable mass value if the second stellar object of the merger of the gravitational wave signal GW190814 would turn out to be a compact star \cite{fattoyev,bombaci2021}), without the need for a velocity of sound close the casual limit but with values, in the most cases, below the conformal limit \cite{traversi}. Neutron stars (actually hadronic stars with hyperonic and $\Delta$ degrees of freedom in the two families scenario) could instead satisfied the constraints obtained from heavy-ion collisions experiments \cite{danielewicz2002} and the tidal deformability constraints derived from GW170817 \cite{fattoyev} which favor softer EOSs. A qualitative agreement of this scenario with the recent NICER results was also showed \cite{traversi,diclemente}.

Concerning the study of the excited dense baryonic matter reachable in heavy-ion collisions, in the seminal work of Ref. \cite{greiner87}, on the basis of the Boguta's $\Delta$ isomers \cite{boguta1982} and in the framework of a non-linear relativistic mean field model, it was predicted that a one-component phase transition from nucleonic matter to $\Delta$-excited nuclear matter can take place in symmetric nuclear matter and the occurrence of this phase transition sensibly depends on the value of the $\Delta$-meson coupling constants. Such a study was also extended with EOSs corresponding to different values of the nucleon effective mass and the saturated compressibility \cite{greiner97}. The range of possible mean-field coupling constants of the scalar and the vector mesons with $\Delta$-isobars, compatible with existence of stable nuclei at the saturation density,  was studied in Ref. \cite{kosov1998}. In a similar framework, the relevance of the $\Delta$-isobar degrees of freedom at different regimes of temperature and density was shown in symmetric and asymmetric hadronic matter with the inclusion of hyperons and the lightest pseudoscalar and vector mesons, by requiring the Gibbs conditions of the global conservation of baryon number, electric charge fraction and zero net strange\-ness \cite{prc2010}.

Following the approach of Ref.s \cite{muller1995,greiner87}, we have studied the presence of thermodynamical instabilities and a subsequent phase transition from nucleonic matter to reso\-nan\-ce-dominated $\Delta$ matter in a warm and dense asymmetric nuclear medium ($T\le 50$ MeV and $\rho_0\le\rho_B\le 3\rho_0$) \cite{prc2012}.

In this paper we plan to extend such previous investigations in regime of high temperature and dense baryon matter with the inclusion of the hyperon and the $\Delta$-isobar degrees of freedom in an effective relativistic hadronic EOS, characterized by a set of mean-field coupling constants compatible with different experimental constraints. By requiring the Gibbs conditions on the global conservation of baryon number and zero net strangeness, we are going to show that the presence of the $\Delta$-isobars can drive to the formation of mechanical (due to fluctuations on the baryon density) and chemical-diffusive instabilities (due to fluctuations on the strangeness density). Analogously to Ref.s \cite{muller1995,prc2012}, an important feature of a system with two conserved charges (baryon number and strangeness content) is that the phase transition is continuous. At variance with the so-called Maxwell construction for one conserved charge, the pressure in not constant in the mixed phase, the binodal coexistence surface is two dimensional and a pure hadronic phase transition with different baryon and strangeness content in the two phases takes place.

\section{Hadronic equation of state}
\label{hadron}

We employ here the scheme of relativistic mean-field (RMF) model at finite temperature and baryon density. For what concern the full octect of the lightest baryons, the dynamics can be described by the following Lagrangian density \cite{glendenning1991,sugahara1994}
\begin{eqnarray}\label{lagrangian}
{\mathcal L}_{\rm octet}&&=
\sum_k\overline{\psi}_k\,[i\,\gamma_{\mu}\,\partial^{\mu}-(M_k-
g_{\sigma k}\,\sigma) -g_{\omega
k}\,\gamma_\mu\,\omega^{\mu}\nonumber\\
&&-g_{\rho k}\,\gamma_{\mu}\,\vec{t}
\cdot \vec{\rho}^{\;\mu}]\,\psi_k +\frac{1}{2}(\partial_{\mu}\sigma\partial^{\mu}\sigma-m_{\sigma}^2\sigma^2)
\nonumber\\
&&-\frac{1}{3}a\,(g_{\sigma N}\,\sigma)^{3}-\frac{1}{4}\,b\,(g_{\sigma N}\,\sigma^{4})+\frac{1}{2}\,m^2_{\omega}\,\omega_{\mu}\omega^{\mu}
\nonumber\\
&&+\frac{1}{4}\,c\,(g_{\omega N}^2\,\omega_\mu\omega^\mu)^2+ \frac{1}{2}\,m^2_{\rho}\,\vec{\rho}_{\mu}\cdot\vec{\rho}^{\;\mu}\nonumber\\
&&-\frac{1}{4}F_{\mu\nu}F^{\mu\nu}-\frac{1}{4}\vec{G}_{\mu\nu}\vec{G}^{\mu\nu} \, ,
\end{eqnarray}
where the sum runs over the full octet of baryons  ($p$, $n$, $\Lambda$, $\Sigma^+$, $\Sigma^0$, $\Sigma^-$, $\Xi^0$, $\Xi^-$) interacting with $\sigma$, $\omega$,
$\rho$ meson fields, $M_k$ is the
vacuum baryon mass of index $k$ and $\vec{t}$ is the isospin operator which acts on the baryon.
The field strength tensors for the vector mesons are given by the usual expressions
$F_{\mu\nu}\equiv\partial_{\mu}\omega_{\nu}-\partial_{\nu}\omega_{\mu}$,
$\vec{G}_{\mu\nu}\equiv\partial_{\mu}\vec{\rho}_{\nu}-\partial_{\nu}\vec{\rho}_{\mu}$.

In the RMF approach, baryons are considered as Dirac quasiparticles moving in classical meson fields and the field operators are replaced by their expectation values. The parameters of the model are fixed to reproduce the properties of equilibrium nuclear matter. In the following we will use the parameters set marked as TM1 of Ref. \cite{sugahara1994}, which has a slightly lower value of the compression modulus $K$ with respect to the GM1 or GM2 sets of Ref. \cite{glendenning1991} and a smaller value of the effective nucleon mass $M^*_N$, more appropriate to reproduce the correct spin-orbit splitting in finite nuclei \cite{serot1998}.

Let us remark that in Ref. \cite{prc2012}, with the same parameters set TM1, we have preliminarily studied the liquid-gas phase transition in re\-gi\-me of low temperature and baryon density for different proton fractions, obtaining results in accordance with previous investigations \cite{muller1995,mekjian2003}.

The meson-hyperon coupling constants has been fix\-ed to the potential depth of hyperons at the saturation density ($U_\Lambda^N=-28$ MeV, $U_\Sigma^N=+30$ MeV, $U_\Xi^N=-18$ MeV) and by means of the SU(6) symmetry relations \cite{schaffner1996,bunta2004}.

We have verified that the two additional meson fields, the hidden strange scalar meson $f_0(975)$ and the vector meson $\phi(1020)$, usually introduced to simulate the hyperon-hyperon attraction observed in $\Lambda-\Lambda$ hypernuclei \cite{schaffner1996,bunta2004}, do not significantly affect the EOS in the considered range of density and temperature and, taking also into account of the uncertainty of the coupling constants, their contributions will be neglected.

On the other hand, as previously discussed, we expect that in regime of finite values of temperature and density, the $\Delta(1232)$-isobar degrees of freedom can play a central role.
To take into account of the $\Delta$-isobars, a formalism was developed considering only the on-shell $\Delta$-particle contribution where the mass of $\Delta$s are substituted by the effective one in RMF approximation \cite{boguta1982,kosov1998,malfliet1992}. In this framework the Lagrangian density for the $\Delta$-isobars can be expressed as
\begin{eqnarray}
{\mathcal L}_\Delta=\overline{\psi}_{\Delta\,\nu}\, [i\gamma_\mu
\partial^\mu -(M_\Delta-g_{\sigma\Delta}
\sigma)-g_{\omega\Delta}\gamma_\mu\omega^\mu
 ]\psi_{\Delta}^{\,\nu} \, ,
\end{eqnarray}
where $\psi_\Delta^\nu$ is the Rarita-Schwinger spinor
for the $\Delta$-isobars.

In literature there are large uncertainties on the couplings $x_{\sigma\Delta}=g_{\sigma\Delta}/g_{\sigma N}$ and $x_{\omega\Delta}=g_{\omega\Delta}/g_{\omega N}$ between $\Delta$s and field mesons  (we limit ourselves to consider only the coupling with the $\sigma$ and $\omega$-meson fields, more of which are explored in the literature, taking also into account of the  high temperature symmetric nuclear matter regime considered in this investigation). Qualitatively, it has been possible to establish that the $\Delta$-isobars inside a nucleus feel an attractive potential \cite{jin1995,marques2018}.
Moreover, as observed in Ref.\cite{prc2014}, from phenomenological analysis of the data relative to electron-nucleus, photoabsorption and pion nucleus scattering can be extracted different experimental constraints on the values of the $\Delta$-meson coupling constants \cite{oset1987,connell1990,plb1994,nakamura2010}.
Of course, the choice of couplings that satisfies the above conditions is not unique but exists a finite range of possible values which depends on the particular EOS under consideration. Without loss of generality, we can limit our investigation by fixing $x_{\omega\Delta}=1$ and varying $x_{\sigma\Delta}$ from unity to the value $x_{\sigma\Delta}=1.2$, compatible with the observational constraints mentioned above. Such values are also consistent with the limits obtained from the data analysis of Ref. \cite{danielewicz2002} (see, for example, Fig. 1 of Ref. \cite{prc2010}). Moreover, we point out that the $\Delta$-metastable condition (appearance of a high density second minimum on the energy per baryon in the zero temperature symmetric EOS), is not realized for the above considered range of couplings. In Ref. \cite{prc2010} a detailed study in absence and in presence of different $\Delta$-meson fields interaction is reported.


The finite temperature and density EOS with respect to strong interaction has to conserve two charges related to baryon number (B) and strangeness number (S). Due to the high temperature involved in this study we will limit to consider symmetric nuclear matter at $Z/A=0.5$ and for simplicity we will not consider fluctuations on the electric charge.
Therefore, the system is described by two independent chemical potentials $\mu_B$, $\mu_S$, and the particle chemical potential of index $i$ can be written as
\begin{equation}
\mu_i=b_i\, \mu_B+s_i\,\mu_S \, , \label{mu}
\end{equation}
where $b_i$ and $s_i$ are the baryon and the strangeness numbers of $i$-th hadronic species, respectively.
On the other hand, the particle chemical potentials are related to the microscopic EOS by means of $\mu_i=\partial\epsilon/\partial\rho_i$ and are given in terms of the effective chemical potentials $\mu_i^*$ as
\begin{equation}
\mu_i=\mu_i^*+g_{\omega i}\,\omega+g_{\rho i}\,t_{3 i}\,\rho\, .
\label{mueff}
\end{equation}
The baryon effective energy is
$E_i^*(k)=\sqrt{k^2+{{M_i}^*}^2}$, where the effective mass of the $i$th baryon is defined as $M_i^*=M_i-g_{\sigma i}\sigma$.

Especially in regime of high temperature and low baryon density, the relevance of the lightest pseudo\-sca\-lar and vector mesons is expected to be important. On the other hand, the contribution of the $\pi$ mesons (and other pseudoscalar and pseudovector fields) vanishes at the mean-field level. From a phenomenological point of view, we can take into account the lightest pseudoscalar ($\pi$, $K$, $\overline{K}$, $\eta$, $\eta'$) and vector mesons ($\rho$, $\omega$, $K^*$, $\overline{K}^*$, $\phi$) as a quasi-particle gas by adding their one-body contribution to the thermodynamical potential (for details, see for example, Refs. \cite{muller1997,prc2010}).

Finally, the thermodynamical quantities can be obtained from the total grand
potential $\Omega$ in the standard way, as a sum of the baryon and meson degrees of freedom.

\section{Phase transitions and stability conditions}

At variance of temperature and density, the multi-com\-po\-nent particles constituent the system can change under the constraint of the global conservation of the baryon number and zero net strangeness.
For such a system, the Helmholtz free energy density $F$ can be written as
\begin{equation}
F(T,\rho_B,\rho_S)= -P(T,\mu_B,\mu_S) +\mu_B\rho_B + \mu_S\rho_S \, ,
\end{equation}
with
\begin{equation}
\mu_B=\left (\frac{\partial F}{\partial\rho_B}\right)_{T,\rho_S} \, , \ \ \ \mu_S=\left ( \frac{\partial F}{\partial\rho_S}\right)_{T,\rho_B} \, .
\end{equation}

By assuming the presence of two phases (denoted as $I$ and $II$, respectively), the system is stable against the se\-pa\-ration in two phases if the free energy of a single phase is lower than the free energy in all two phases configuration. In this case the phase coexistence is described by the following Gibbs conditions
\begin{eqnarray}
&&\mu_B^{I} = \mu_B^{II} \, , \ \ \ \ \ \ \ \ \ \mu_S^{I} = \mu_S^{II}
\, , \label{gibbs1}\\
&&P^I (T,\mu_B,\mu_S)=P^{II} (T,\mu_B,\mu_S) \, .
\label{gibbs2}
\end{eqnarray}
At a given baryon density $\rho_B$ and at a zero net stran\-ge\-ness density ($r_S=\rho_S/\rho_B=0$), the chemical potentials $\mu_B$ and $\mu_S$ are univocally determined by the following equations
\begin{eqnarray}
&&\!\!\rho_B=(1-\chi)\,\rho_B^I(T,\mu_B,\mu_S) +\chi \,\rho_B^{II}(T,\mu_B,\mu_S) \, ,\label{rhobchi}\\
&&\!\!\rho_S=(1-\chi)\,\rho_S^I(T,\mu_B,\mu_S) +\chi \,\rho_S^{II}(T,\mu_B,\mu_S) \, ,
\label{rhoschi}\end{eqnarray}
where $\rho_B^{I(II)}$ and $\rho_S^{I(II)}$ are, respectively, the baryon and strangeness charge densities in the lower density ($I$) and in the higher density ($II$) phase and $\chi$ is the volume fraction of the phase $II$ in the mixed phase ($0\le\chi\le 1$).

Unlike the case of a single conserved charge, where the pressure in the so-called Maxwell construction is constant, for two conserved charges the pressure in the mixed phase is not constant and the baryon and the strangeness densities can be locally different in the two phases, although the total $\rho_B$ and $\rho_S$ of system result to be globally conserved. At the thermal equilibrium, the possible phase transition can be characterized by mechanical (fluctuations on the baryon density) and chemical instabilities (fluctuations on the strangeness density) with a consequent two dimensional binodal coexistence surface \cite{das2015,gulminelli2016,das2017,gulminelli2017}.

The condition of the mechanical stability implies \cite{muller1995}
\begin{equation}
\rho_B \left(\frac{\partial P}{\partial \rho_B}\right)_{T,\,\rho_S} >0   \, ,  \label{InstabMecc}
\end{equation}
therefore, when the compressibility becomes negative, at fixed temperature and strangeness density, a mechanical instability appears  in the EOS.

By defining  $\mu_{i,j}=(\partial\mu_i/\partial\rho_j)_{T,P}$ (with $i,j=B,S$) \cite{reichl}, the chemical stability can be expressed with the following conditions
\begin{eqnarray}
\mu_{B,B} >0 \, , \ \ \ \mu_{S,S}>0 \, , \ \ \
\begin{vmatrix}
\,\mu_{B,B} & \mu_{B,S} \\
\,\mu_{S,B} & \mu_{S,S}
\end{vmatrix}
>0  \,  . \label{InstabChim}
\end{eqnarray}
In addition to the above conditions, for a process at constant $P$ and $T$, it is always satisfied that
\begin{eqnarray}
&&\rho_B\, \mu_{B,B}+\rho_S \, \mu_{S,B}=0\, , \\
&&\rho_B\, \mu_{B,S}+\rho_S \, \mu_{S,S}=0\, . \label{diff1}
\end{eqnarray}

More explicitly, for example, Eq.(\ref{diff1}) can be written as
\begin{equation}
\left(\frac{\partial \mu_B}{\partial r_S}\right)_{T,P}+r_S\, \left(\frac{\partial \mu_S}{\partial r_S}\right)_{T,P}=0 \, .\label{diff_delta}
\end{equation}

The system has a zero net strangeness content but during a phase transition the strangeness fraction $r_S$ is not locally fixed in the single phase. At a given  temperature, during the compression of the system, the appearance of strange particles/antiparticle could, in principle, shift the diffusive instability region to positive or negative values of $r_S$. Such a feature has no counterpart in the standard liquid-gas phase transition where the proton fraction is always positive \cite{muller1995}.

Taking into account of these aspects, the chemical stability condition is satisfied if
\begin{equation}
\!\!\left(\frac{\partial \mu_S}{\partial r_S}\right)_{T,P}>0 \ \ {\rm or } \ \ \left\{
\begin{array}{rl}
\displaystyle
\left(\frac{\partial \mu_B}{\partial r_S}\right)_{T,P}<0\,, & \mbox{if } r_S>0 \, , \\
\\
\displaystyle
\left(\frac{\partial \mu_B}{\partial r_S}\right)_{T,P}>0\,, & \mbox{if } r_S<0 \, .
\end{array}
\right.
\end{equation}

Whenever the above stability conditions are not respected, the system becomes unstable and a binodal surface in $(T,P,r_S)$ space encloses the area where the system undergoes to the phase transition.

\section{Results and discussion}\label{result}

We are now able to investigate the presence of thermodynamic instabilities in the symmetric nuclear EOS at different values of temperature and baryon density.

As already anticipated, the presence of the $\Delta$-isobar degrees of freedom plays a crucial role into the formation of thermodynamic instabilities. Although unstable conditions can be realized for different combinations of the meson-$\Delta$ coupling constants, corresponding to a larger net attraction for $\Delta$ isobars with respect the nucleon one, we initially focalize our discussion by fixing $x_{\sigma \Delta}=1.2$ and $x_{\omega \Delta}=1$, values compatible with different experimental constraints previously discussed \cite{prc2014}. Let us observe that, in the case of a net repulsive $\Delta$-interaction or in absence of interaction, the effects thermodynamic instabi\-li\-ties would disappear or become negligible. This is mainly due to the softening of the EOS with the appearance of $\Delta$ isobars, which favor, together with hyperons, the formation of mechanical instabilities (\ref{InstabMecc}). On the other hand, an attractive $\Delta$-interaction modifies, at fixed $\mu_B$ and finite $T$, the strange chemical potential $\mu_S$ (see for example, Fig. 9 of Ref. \cite{prc2010}) by affecting the presence of chemical (strangeness) instabilities.

\begin{figure}[h]
\begin{center}
\resizebox{0.48 \textwidth}{!}{%
\includegraphics{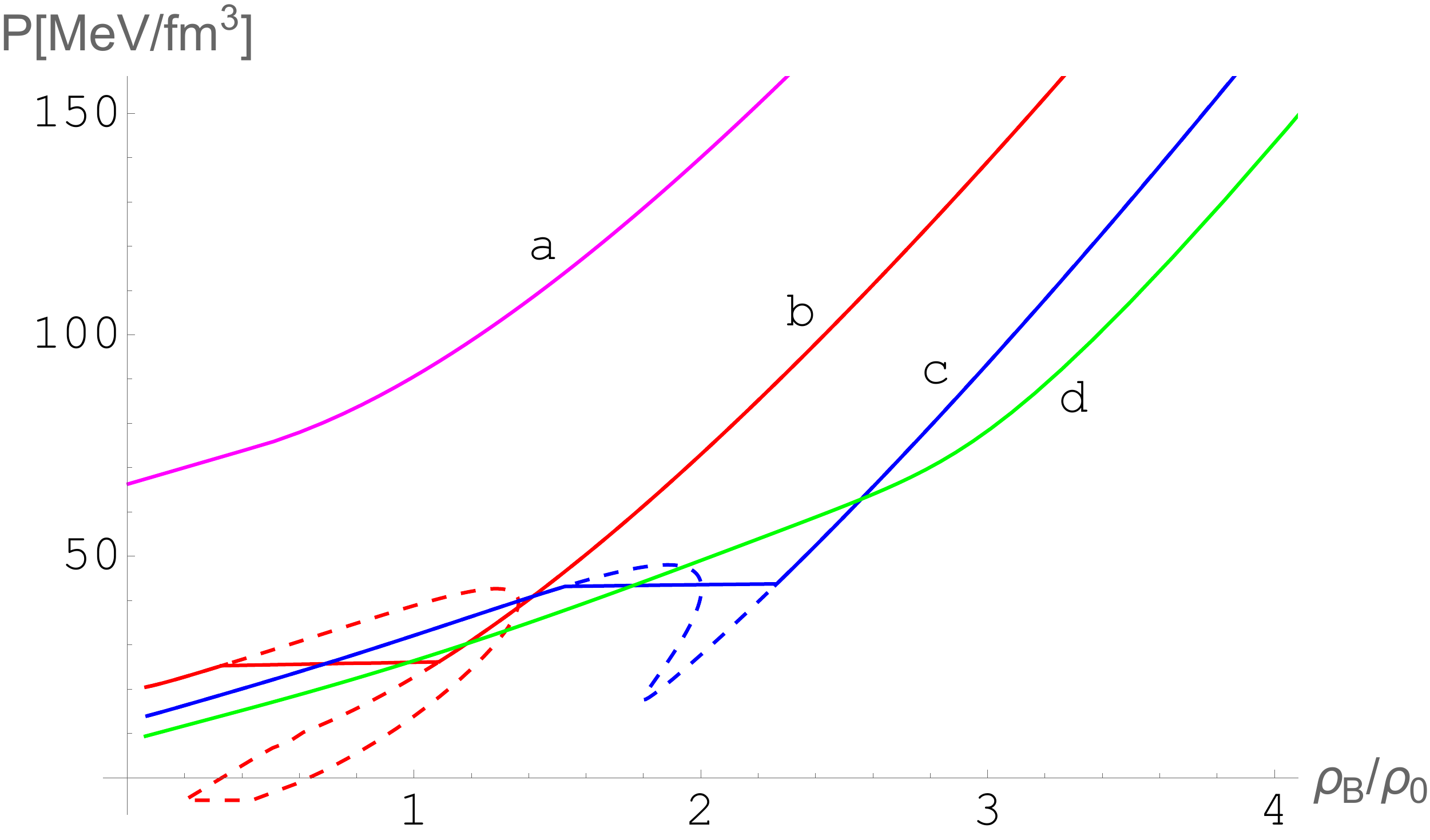}
} \caption{Pressure as a function of the baryon density (in units of the nuclear saturation density $\rho_0$) at different temperatures. The curves labeled $a$ through $d$ have decreasing temperatures: $T$= 150, 140, 130 and 120 MeV, respectively. In the case $b$ ($T$=140 MeV) and $c$ ($T$=130 MeV), the system is mechanically unstable and the continuous (dashed) lines correspond to the solution obtained with (without) the Gibbs construction.} \label{fig_prhob}
\end{center}
\end{figure}

In Fig. \ref{fig_prhob}, we show the pressure as a function of the baryon density at different temperatures and zero net stran\-geness ($r_S=0$). For the curves $b$ (corresponding to $T=140$ MeV) and $c$ (corresponding to $T=130$ MeV), the condition (\ref{InstabMecc}) is clearly not satisfied and the mechanical instabilities are realized from about $T=125$ MeV to $T=145$ MeV, over a finite range of baryon densities. For the unstable isotherms, $b$ ($T$=140 MeV) and $c$ ($T$=130 MeV), the continuous lines correspond to the solution obtained with the Gibbs construction, related to the conditions (\ref{gibbs1}) and (\ref{gibbs2}), whereas the (unphysical) dashed lines with the appearance of loops are without correction.

\begin{figure}[h]
\begin{center}
\resizebox{0.48 \textwidth}{!}{%
\includegraphics{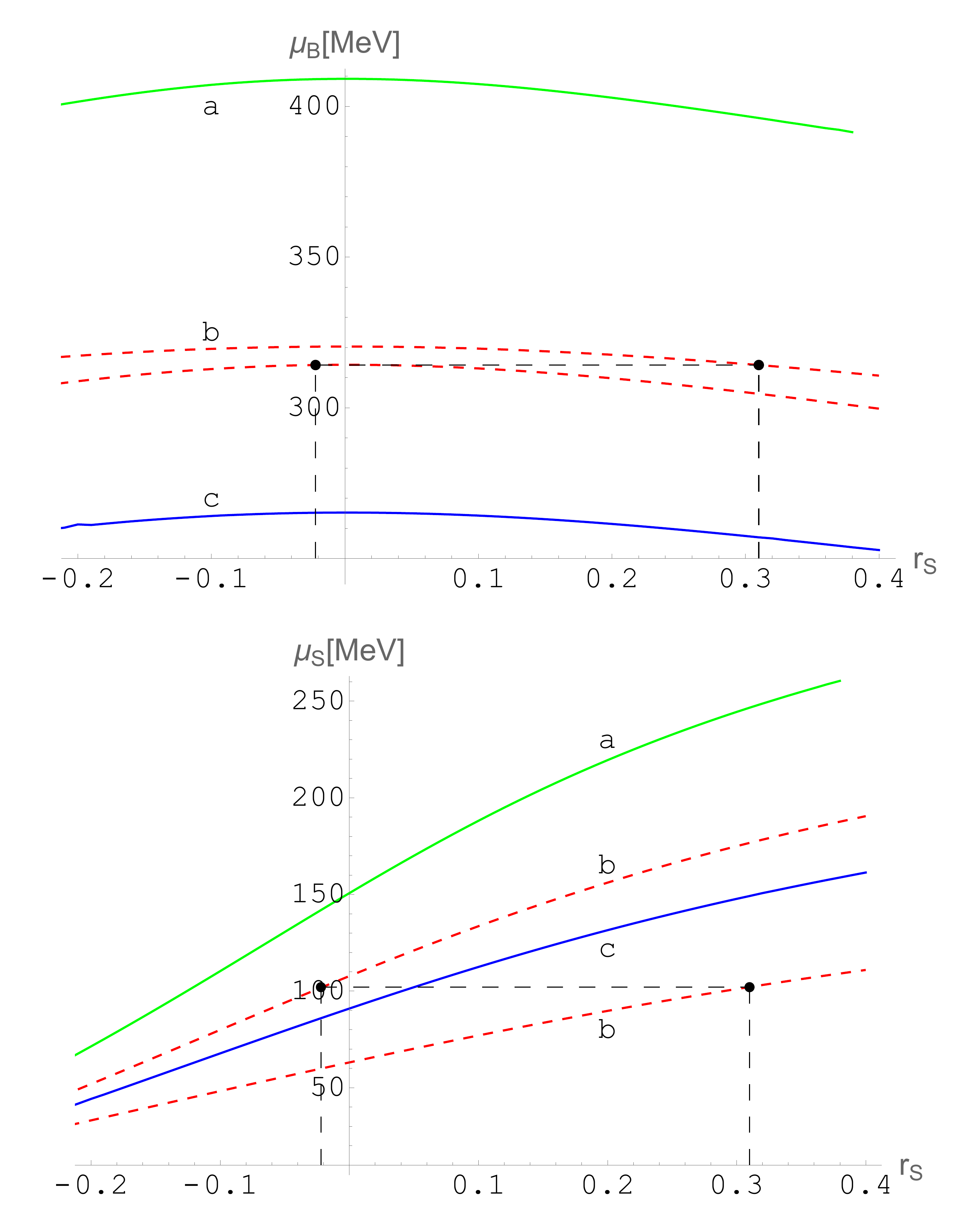}
} \caption{Baryon (upper panel) and strangeness (botton panel) chemical potentials at $T=140$ MeV as a function of the strangeness ratio $r_S$. The curves labeled $a$, $b$ and $c$ correspond to a value of pressure  $P$=50, 26 and 20 MeV/fm$^3$, respectively. In the case $b$ ($P$=26 MeV/fm$^3$), the system results to be unstable and the geometrical construction of the Gibbs conditions is reported in the rectangular region.} \label{rs_mui}
\end{center}
\end{figure}

In the most cases, together with the presence of the mechanical instability, the chemical instability conditions result to be also achieved. To better clarify the realization of this last condition, in Fig. \ref{rs_mui}, we report the two independent chemical potentials $\mu_B$ (upper panel) and $\mu_S$ (lower panel) for  three different values of pressure ($P$=20, 26 and 50 MeV/fm$^3$) as a function of the strangeness fraction $r_S$, at $T$=140 MeV.

The cases $a$ and $c$ correspond to a value of pressure for which the chemical stability conditions are satisfied and, at fixed value of $r_S$, we have a unique value of $\mu_B$ and $\mu_S$. Otherwise, the red dashed lines, labeled with $b$ in the two panels, show an example of chemical instability due to a multiple solution for the chemical potentials at fixed value of pressure. In the black points at the edges of the rectangular regions are reported the geometrical constructions of the phase equilibrium, on the basis of the Gibbs conditions: the pressure and the chemical potentials of the two phases at different strangeness fractions, $r_S^{(1)}$ and $r_S^{(2)}$, are equal at the same temperature. The collection of all pairs of strangeness fractions $r_S^{(1)}(T,P)$ and $r_S^{(2)}(T,P)$, defines the binodal surface, which encloses the area of thermodynamical instability of the system.

\begin{figure}
\begin{center}
\resizebox{0.48 \textwidth}{!}{%
\includegraphics{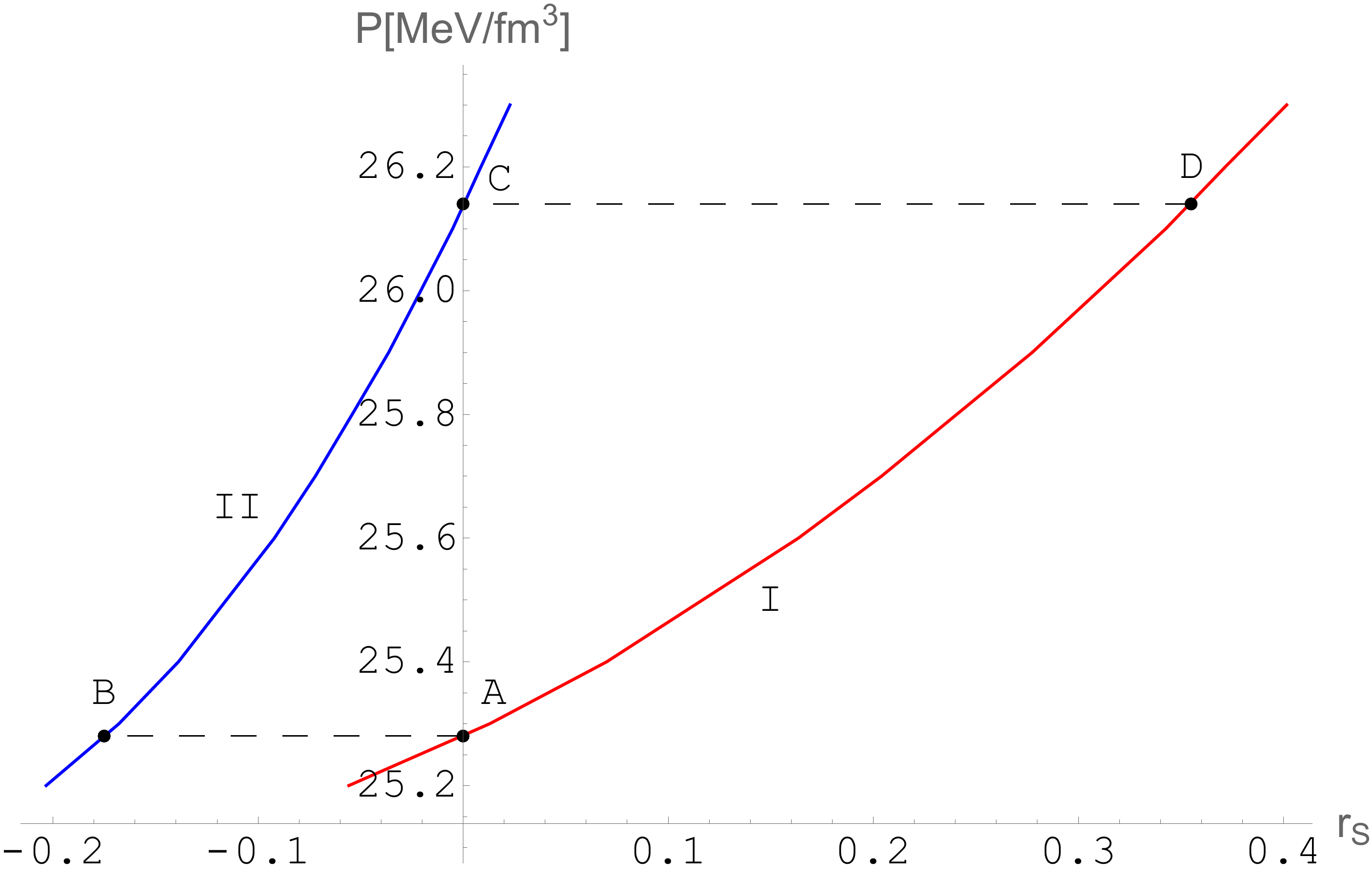}
} \caption{Binodal section giving the two phase coexistence phase boundary in the mixed phase at $T$=140 MeV.} \label{Dia_Binod}
\end{center}
\end{figure}
In Fig \ref{Dia_Binod}, we report the corresponding binodal section at $T$=140 MeV.
During the isothermal compression, the system meets the unstable region in the point $A$, at  $\rho_B(A)\approx 0.3\,\rho_0$ and $r_S=0$, and it separates into two phases of different strangeness ratio $r_S$. At the same time, a second phase appears in $B$ at higher baryon density, $\rho_B(B)\approx 1.1\,\rho_0$. Then each phase evolves from $A$ to $D$ (phase $I$) and from $B$ to $C$ (phase $II$) with an almost constant baryon density in each phase. Finally, the system emerges in the higher density phase in $C$, at the same strangeness fraction of $A$ ($r_S=0$). Let us observe that the phase transition occurs in a very strictly range of pressure corresponding ho\-we\-ver to a sensible variation in the baryon density (about $0.8\, \rho_0$) at a baryon chemical potential $\mu_B\simeq$ 320 MeV (in the case of $T$=130 MeV, the phase transition occurs at about $\mu_B\simeq$ 600 MeV, with an almost constant baryon density of about $1.5\,\rho_0$ in the phase $I$ and $2.2\,\rho_0$ in the phase $II$).

Therefore, in the mixed phase, two phases at different baryon density and strangeness content take place. The phase $I$, at lower density and positive strangeness with an excess of $\overline{s}$ quarks, corresponding to an enhancement of anti-hyper\-ons and $K^+$, $K^0$ mesons. As a counterpart, the phase $II$, at higher density and negative strange\-ness with an excess of $s$ quarks, due to the formation of hyperons and $K^-$, $\overline{K}^0$ mesons (in addition to a $\Delta$-rich matter). This feature has strictly analogies to the quark-hadron phase transition where it possible to realize the so-called strange\-ness distillation: $\overline{s}$ quarks are foreseen mainly present in the lower density hadronic phase and the population of $s$ quarks should be greatly enriched in the higher density quark-gluon phase \cite{greiner1987prl,greiner1991prd,jpg2012}.

\begin{figure}[h]
\begin{center}
\resizebox{0.48 \textwidth}{!}{%
\includegraphics{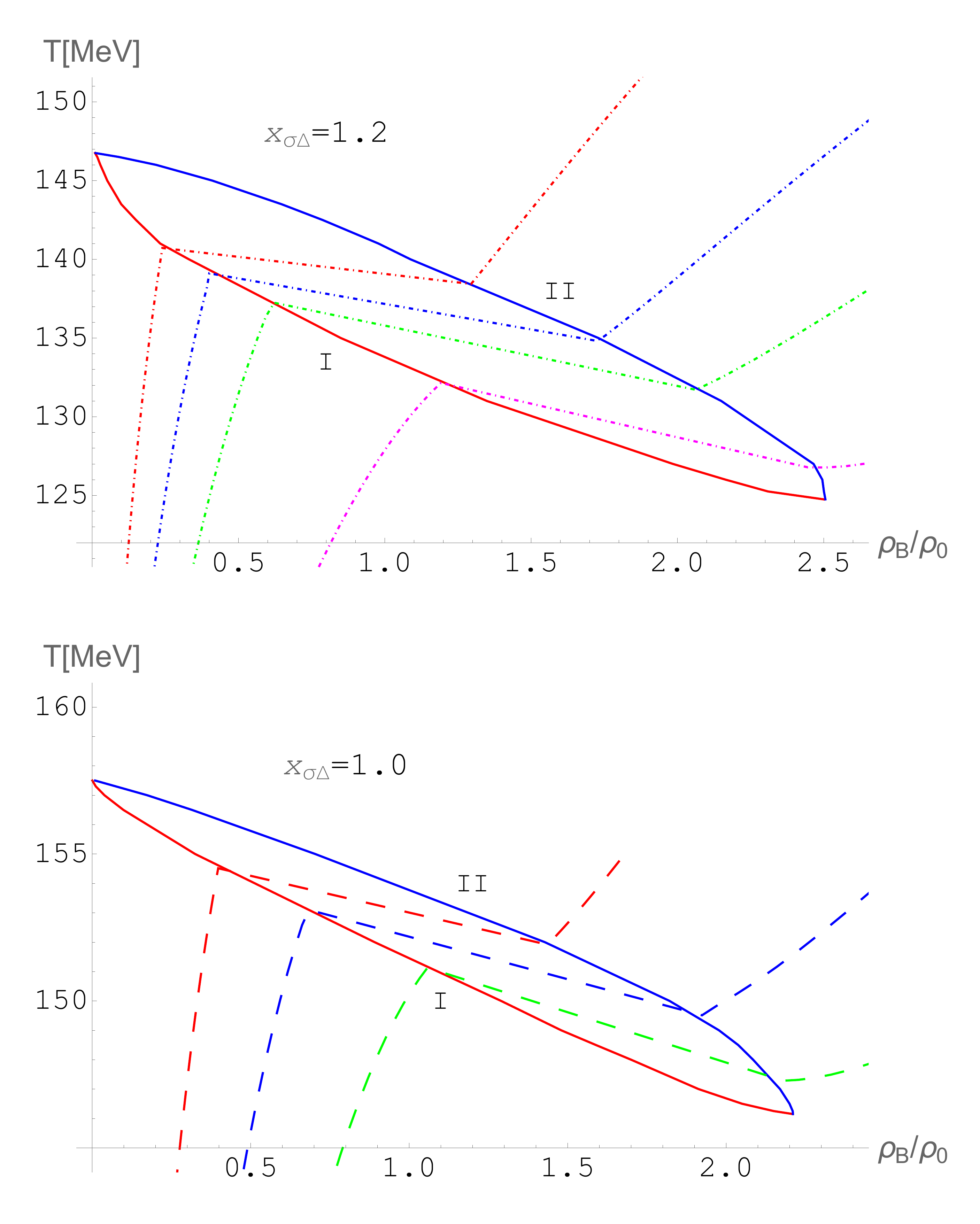}
} \caption{Phase diagrams for two values of the coupling: $x_{\sigma \Delta}=1.2$ (upper panel) and $x_{\sigma \Delta}=1.0$ (lower panel). Dot-dashed and dashed lines, represent the isentropic trajectories for $S/B=30, 20, 15, 10$ (red, blue, green and magenta, respectively) for the two coupling constants $x_{\sigma \Delta}$.} \label{Dia_Fase}
\end{center}
\end{figure}

As previously outlined, the region in which the thermodynamic instabilities take place is very sensitive to the value of the  $x_{\sigma \Delta}$ coupling constant.
At this regards, in Fig. \ref{Dia_Fase}, we report the phase diagram in the temperature-baryon density plane, at a fixed value of $x_{\omega\Delta}=1$ and for the values $x_{\sigma \Delta}=1.2$ (upper panel) and $x_{\sigma \Delta}=1$ (lower panel). Different isentropic lines corresponding to the values $S/B=30, 20, 15, 10$ (red, blue, green and magenta, respectively) are also reported.

Let us observe that the thermodynamic instabilities are already present in the so-called "minimal coupling" choice, assuming the $\Delta$-isobars coupling constants equal to the nucleon one ($x_{\sigma \Delta}=x_{\omega\Delta}=1$). By increasing $x_{\sigma \Delta}$ and, consequently, the relevance of the $\Delta$-isobar degrees of freedom in the EOS, we observe a remarkable reduction of the critical temperature and an increase of the baryon density range for which the system enters into the thermodynamical instabilities region. Furthermore, along each isentropic trajectory, conserved in a fluid element in the hydrodynamics mo\-dels \cite{nonaka2005}, we have in the mixed phase a reduction of the temperature in a wide range of baryon density. This peculiar behavior could be phenomenologically relevant in order to identify such a phase transition in the future compressed baryonic matter experiments \cite{cmb2021,jparc2021,shine2020,bleicher2020,star2017}.

We have verified that the baryon effective masses never become negative in the range of the considered coupling constants. In presence of the phase transition, we observe a remarkable reduction of the effective masses during the mixed phase. This effect is relatively stronger for the nucleons and $\Delta$ isobars. For example, at $T=$140 MeV, with $x_{\sigma \Delta}$=1.2, the nucleon ratio $M^*/M_N$ is reduced to 0.08, while the $\Delta$ isobars ratio $M^*/M_\Delta\simeq 0.13$ at the end of the phase transition, corresponding to $\rho_B\simeq 1.1\,\rho_0$. At densities greater than the second transition density, the effective masses decrease very slowly.
In this context, let us observe that such effective masses cannot be directly compared with the baryon ground state masses obtained in lattice QCD predictions at vanishing baryon density \cite{aarts}.

Finally, concerning Fig. 4, it is necessary to observe that, in order to complete the phase diagram, we have extended our results to very low baryon densities even if the considered EOS is mainly appropriate at finite baryon density.

\begin{figure}[h]
\begin{center}
\resizebox{0.48 \textwidth}{!}{%
\includegraphics{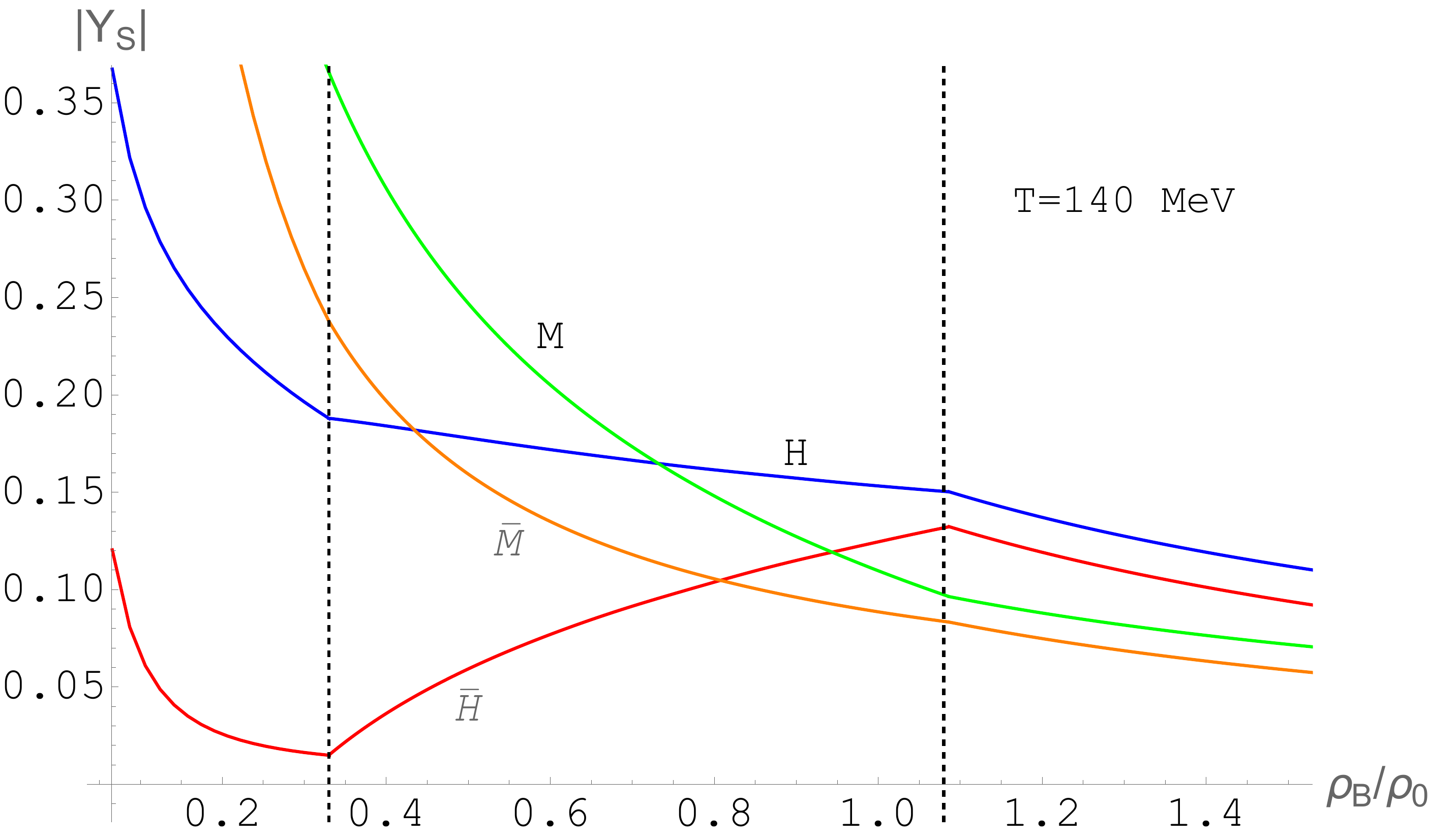}
}
\caption{Absolute value of the strangeness fractions $Y_S=\rho_S/\rho_B$ for hyperons ($H$), anti-hyperons ($\overline{H}$), strange mesons ($M$) and strange anti-mesons ($\overline M$) as a function of the net baryon density at $T=140$ MeV and $x_{\sigma \Delta}=1.2$. The vertical dashed lines delimit the regions of the mixed phase.} \label{Rtot}
\end{center}
\end{figure}

In order to get a deeper insight into the chemical particle composition during the phase transition, in the Fig. \ref{Rtot}, we report the absolute value of the stran\-ge\-ness fractions $Y_S=\rho_S/\rho_B$ for hyperons, anti-hyperons, strange mesons and anti-me\-sons as a function of the net baryon density at $T=140$ MeV and $x_{\sigma \Delta}=1.2$. In accordance with the comments of Fig. \ref{Dia_Binod}, during the mixed phase we have a strong enhancement of anti-hyperons mainly in the lower density mixed phase $I$ with positive strangeness. The global zero net strange\-ness is realized by means of a slo\-wer reduction of the strange anti-mesons (ma\-in\-ly present in the higher density mixed phase $II$) with respect to the strange mesons. At the end of the phase transition (at about $\approx 1.1\,\rho_0$), the strangeness fraction decreases with approximately the same slope for strange baryons and mesons.



%
\begin{figure}
\begin{center}
\vspace{0.4cm}
\resizebox{0.48 \textwidth}{!}{%
\includegraphics{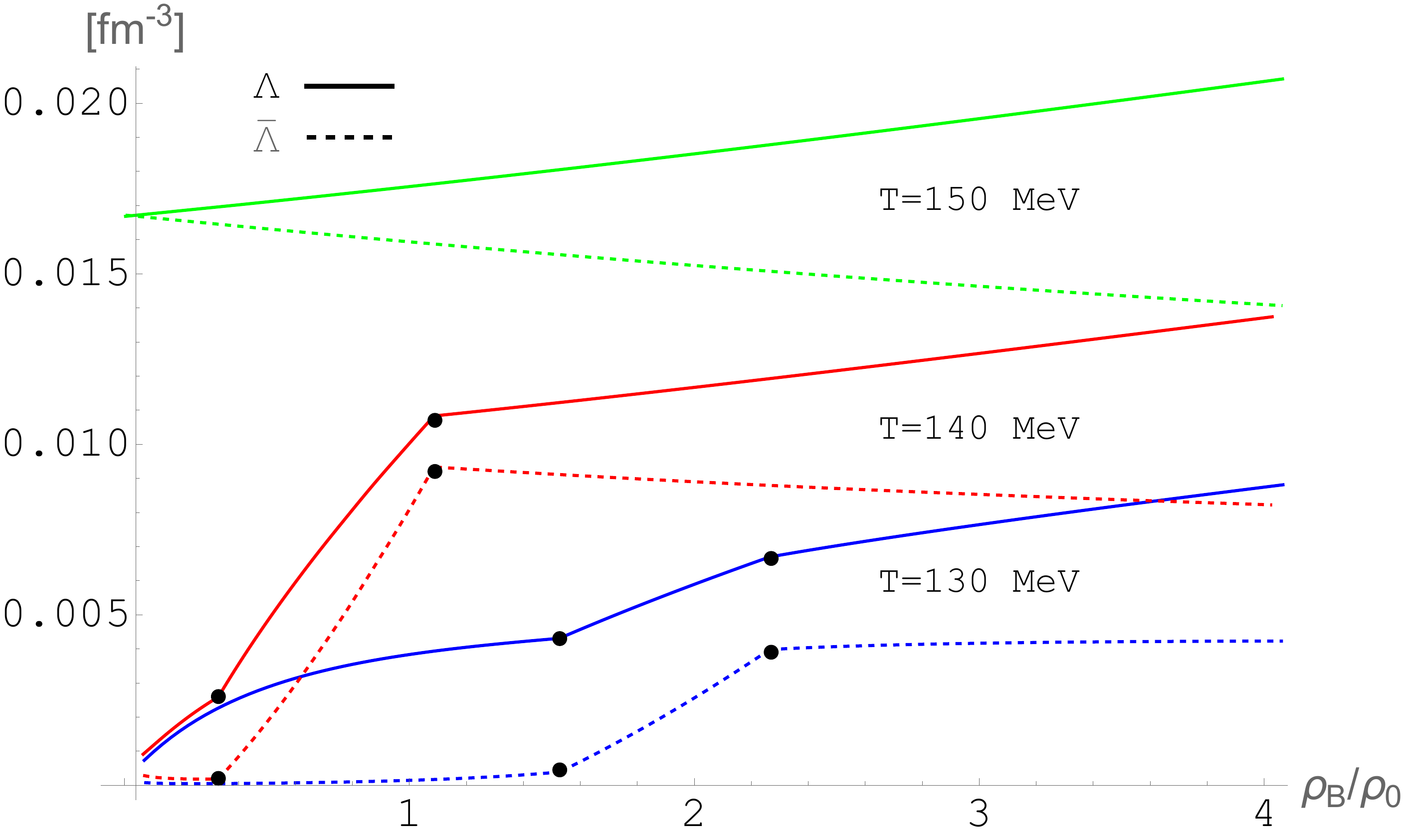}
}
\caption{$\Lambda$ (solid lines) and $\overline{\Lambda}$ (dashed lines) densities (in units of fm$^{-3}$) as a function of the net baryon density for diffe\-rent temperatures. Dots delimit the mixed phase region where thermodynamic instabilities are present ($T=140$ MeV, red lines and $T=130$ MeV, blue lines).}
\label{KpKm}
\end{center}
\end{figure}
In Fig. \ref{KpKm}, we report the $\Lambda$ (solid lines) and $\overline{\Lambda}$ (dashed lines) particle densities as a function of the baryon density, for different temperatures and  $x_{\sigma \Delta}=1.2$. The dots delimit the region of the mixed phase at $T=130$ (blue lines) and $140$ MeV (red lines), where thermodynamic instabilities are present (the system becomes unstable for $T\gtrsim 125$ MeV). According to the previous discussion, by increasing the baryon density during the mixed phase, we have an enhancement for both $\Lambda$ (mainly in the higher density phase $II$) and $\overline{\Lambda}$ (mainly in the lower density phase $I$) but this effect is stronger for $\overline{\Lambda}$. We get a similar behavior also for the other strange bary\-ons, even if with lower particle densities (in comparison, let us observe that in Fig. \ref{Rtot} the strangeness densities have been divided by the baryon density). As a counterpart, we have found that a sharp reduction in the strange meson/anti-meson ratios (mainly in the $K^+/K^-$ ratio) occurs into  the mixed phase.





\section{Conclusions}


Nuclear phase transitions and critical phenomena have been studied at different regimes of temperature and baryon density reachable in relativistic heavy-ion collisions. High energy compressed baryonic matter experiments will open the possibility to investigate in detail finite temperature and dense nuclear matter.

The main goal of this work it to show the possible pre\-sen\-ce of thermodynamical instabilities at high temperature and dense nuclear matter, by requiring the global conservation of the baryon number and zero net strangeness. Similarly to the liquid-gas phase transition in asymmetric nuclear matter, mechanical and chemical-diffusive thermodynamic instabilities can be formed but, in the present regime, the corresponding phase transition is driven by a different strangeness content in the mixed phase, instead of a different electric charge fraction.

The considered effective EOS has the noticeable advantage of making the non trivial numerical analysis more easy to handle, even if cannot, of course, to incorporate the complex many-body interactions at finite temperature and baryon density. It would be very interesting to extend such a study to a more realistic chiral symmetric model and beyond of the mean field approximation.

As first observed in Ref. \cite{greiner87}, the introduction of the $\Delta$ isobar degrees of freedom plays a crucial role in the realization of the unstable conditions, which are sensible to the values of the $\Delta$-meson coupling constants. We have seen that the mechanical and the chemical thermodynamic instabilities appear in the EOS considering a finite range of couplings compatible with different expe\-ri\-mental constraints.

Differently from the discontinuous one-com\-po\-nent pha\-se transitions, for a two-component system a continuous ha\-dro\-nic phase transition takes place with two pha\-ses at the same baryon and strangeness chemical potentials but with a different content of baryon and strangeness density. A phase $I$, at lower baryon density and positive strange\-ness and a phase $II$, at higher baryon density, negative strange\-ness and $\Delta$-rich matter.

Due to the global conservation of zero net strange\-ness, during the phase transition, at fixed temperature, we observe a pure hadronic strange\-ness distillation, a strong enhancement of the anti-hyperon to hyperon ratios with a consequent formation of $\overline{s}$ quarks, mainly in the baryon sector in the lower density phase $I$, and of $s$ quarks, mainly in the meson sector in the higher density phase $II$.
Furthermore, the considered hadronic phase transition, which implies a softening of the EOS, have very similar features and signatures to the hadron-quark phase transition with an analogue strangeness distillation effect due to a large anti-strangeness content in the hadron phase while the quark-gluon phase retains a large net strangeness excess \cite{greiner1987prl,greiner1991prd}.
In this context, let us observe that the formation of a high density $\Delta$-rich matter in the hadronic phase can delay the hadron-quark phase transition at fixed temperature \cite{jpg2012}.


In the last years, many important progresses have been made in the theoretical modeling of high baryon density with the development of hydrodynamic and microscopic transport models to simulate space-time evolution of hot and dense nuclear matter generated in high energy heavy-ion collisions \cite{sorge1989,aichelin1991,buss2012,koncha2014,ivanov2016a,batyuk2016,nara2016,nara2018,nara2019,shen2020}. Analysis of collective flows, such as directed and elliptic flow, which are sensitive to the early stage of the collisions, can give valuable information about the nuclear EOS \cite{stoecker2005,ivanov2016b,nara2020}.

However, to date, the developed hydrodynamic and transport models seem to have been unsuccessful in the reproducing the beam energy dependence of the directed flow slope within a single EOS parameters set \cite{koncha2014,batyuk2016,ivanov2016b,nara2020}. In particular, the NA49 Collaboration \cite{na49} and, mo\-re recently, with a much higher statistics, the STAR Collaboration \cite{star1,star2} clearly discovered a change of sign of the proton directed flow slope around $\sqrt{s_{NN}}=10$ GeV at mid-rapidity. On one side, quantum molecular dynamic transport models well reproduces the experimental directed and elliptic flow by means of a rather stiff mono\-to\-nous EOS up to $\sqrt{s_{NN}}=8.8$ GeV, whereas the collapse of the proton direct flow at higher energy beam seems to support a softening of the EOS around $\sqrt{s_{NN}}=10$ GeV, corresponding to a (unknown) first order phase transition \cite {nara2019,nara2020} \footnote{ Within three-fluid dynamics simulations, a smooth crossover seems instead to be favored by the most part of considered experimental data \cite{ivanov2016a,ivanov2016b}.}. In this context, we observe that most of the theoretical calculations predict the collapse of the directed flow below $\sqrt{s_{NN}}\approx 6$ GeV.

It is still premature to conclude unambiguously that the collapse of the directed flow is a clear signature of a phase transition, on the other hand the hypothetical softening of the EOS could be in principle compatible with the pure hadronic phase transition of the present investigation (also due to the similarities with a hadron-quark phase transition). Although the results of hydrodyna\-mic and microscopic hadron transport models are very sensitive to the considered assumptions and the adopted EOS, the order of magnitude of different physical quantities that characterize the phase transition (such as the values of entropy per net baryon $S/B\approx$ 18$\div$25, temperature and baryon densities involved in the dynamical trajectories at $\sqrt{s_{NN}}=$7.7 and 11.5 GeV, predicted in Ref. \cite{ivanov2016a}; the values $S/B=$ 10, 20 and the pressure during the first order phase transition considered in Ref.s \cite{nara2016,nara2018}), appears to be comparable with that involved in the thermodynamic instabilities region here consi\-de\-red.

Among the others, detailed and simultaneous studies of the radial, directed, elliptic flow values \cite{nara2018} and/or sophisticated analysis, such as the extraction of the bulk modulus \cite{prl2007}, could discriminate more clearly the occurrence of a pure hadronic or the nature of a hadron-quark phase transition in the compressed baryon matter regime.


\vspace{0.5cm}
\noindent
{\bf Acknowledgments}\\
It is a pleasure to thank A. Drago and G. Pagliara for useful discussions.

\end{document}